\documentstyle[prl,aps,graphicx,twocolumn]{revtex}

\newcommand{\beq}{\begin{equation}}
\newcommand{\eneq}{\end{equation}}

\begin{document}

\begin{center}
{\bf {\large On attraction between particles with fractionalized quantum
numbers in one-dimensional devices}}

\vspace*{0.5cm} {\sl Bogdan A. Bernevig}$^*$,
{\sl Domenico Giuliano}$^{\dagger }$ and {\sl Robert B. Laughlin} $^{ * }$

\bigskip {\small $^{\dagger }$ Dipartimento di Fisica, Universit\`{a} della
Calabria and \\
I.N.F.N., Gruppo collegato di Cosenza,
Arcavacata di Rende - Cosenza, Italy} \\
 {\small $^{* }$ Department of Physics, Stanford University, Stanford
(CA) 94305}

\bigskip
\end{center}

\begin{quotation}
We make some comments about the results we obtained in
\cite{us1,us2} and show that the conclusion of a recent paper
\cite{greiter} leveling some criticism on our results is, in fact,
incorrect.
\end{quotation}

\bigskip 

\vspace*{0.3cm}

Recently, a note has appeared in the cond-mat archive
\cite{greiter}, disagreeing with the interpretation of the results
obtained in \cite{us1,us2}.  These results concern the attraction
between particles with fractionalized quantum numbers in
one-dimensional systems. The note \cite{greiter} convinced us that
it is worth providing some side comments on the physics discovered
in \cite{us1,us2}. In particular, we shall focus our attention onto
the case of ``half spin-waves'' (spinons) in the Haldane-Shastry
(HS)-model \cite{duncan}.

Since several authors referred to HS-Hamiltonian as an ideal gas of
noninteracting semions, this way of looking at a spinon gas
seems not to be consistent with our results, showing that there is, in fact,
a short-range attraction between spinons. To complement the various comments
we made on such a point in Ref.\cite{us1}, we would like to stress that it
is, in fact, true that the $S$-matrix for spinon scattering does not
depend on the momentum, if one labels spinon excitations with their
quasimomenta \cite{essler}. On the other hand, we start with the
wavefunction for two localized spinons, $\Psi_{\alpha \beta}$
\cite{us1}. We define the two-spinon wavefunction by decomposing
 $\Psi_{\alpha \beta}$ in the basis of energy eigenstates, $\Phi_{mn}$, which
is not overcomplete. Contrary to remarks in \cite{greiter}, our
definition is, therefore, not affected by any ambiguity. In fact,
this can be seen from the characteristic equation for the localized
two-spinon wavefunction $\Psi_{\alpha \beta}$, with which
\cite{greiter} agrees:

{\begin{displaymath} \biggl[- \left(z_\alpha \frac{\partial}{\partial
z_\alpha} \right)^2 - \left(z_\beta \frac{\partial}{\partial
z_\beta} \right)^2 +
\end{displaymath}

\begin{displaymath}
+ (\frac{N}{2} -1) \left( z_\alpha \frac{\partial}{\partial
z_\alpha} +z_\beta \frac{\partial}{\partial z_\beta} \right)  + (2
M^2 -M(N-2)) -
\end{displaymath}
\begin{equation}
- \frac{1}{2}\frac{z_\alpha + z_\beta}{z_\alpha - z_\beta} \left(
z_\alpha \frac{\partial}{\partial z_\alpha} - z_\beta
\frac{\partial}{\partial z_\beta} \right)\biggr] \Phi_{\alpha \beta}
= \lambda \Phi_{\alpha \beta}
\end{equation}}

\noindent where $z_\alpha, z_\beta$ are the positions of the two
spinons, $N$ is the number of particles in the chain, and $M=N/2
-1$. The momentum and position dependent spinon interaction is
obviously manifest in the last term of the equation. The apparent
contradiction with the result in \cite{essler} is just a consequence
of a different way of labelling spinons. Quasimomenta are good
quantum numbers when exactly solving correlated Hamiltonians but,
unfortunately, they are not observable. Within our representation,
we show a different aspect of the problem, that is, that spinon
interaction enhances the probability for two spinons to be at the
same point, which is the quantity that appears in the finite-chain
version of Haldane-Zirnbauer formula for the dynamical spin
susceptibility \cite{hz}. Within the quasi-momenta representation
chosen by \cite{greiter}, the energy of the interaction disappears,
but the spinons close the gap at different values of their momenta.
The re-scaling of an interaction energy in the kinetic energy of
spinons can always be done, but it is just an artifact and it is not
physical. The interaction is then hidden in the difference of
momenta ($m,n$) between spinons at the zero-gap point.

In conclusion, we believe it is clear that our results do not
contradict elegant solutions of the HS-model, like in
Ref.\cite{essler}, but that they rather complement them, as widely
discussed in Ref.\cite{us1}. Moreover, our derivation does not
suffer from the ambiguity of overcompleteness, and clearly shows
that spinons interact through \emph{short-range} forces.

\vspace{.5cm}

We would like to thank D. I. Santiago, A. Tagliacozzo, and M. Berg\'ere for
useful discussions.

\end{document}